\documentclass[a4paper]{jpconf}
\usepackage{slashed}
\usepackage[latin1]{inputenc}
\usepackage{times}
\usepackage{amsmath}
\usepackage{amssymb}
\usepackage{amsthm}
\usepackage{verbatim}
\usepackage{graphicx}
\usepackage{color}
\usepackage{booktabs}
\usepackage{subfigure}
\newcommand{\dint}{{\displaystyle \int}} 
\newcommand{\nths}{\negthickspace\negthickspace}

\begin{document}
\title{On the equivalence of semi-classical methods for QED in intense external fields}

\author{Anthony Hartin}

\address{DESY FLC, Notkestrasse 85, Hamburg, Germany}

\ead{anthony.hartin@desy.de}

\begin{abstract}
Using the semi-classical method of Nikishov-Ritus (NR), the derivation of the transition rate of the beamsstrahlung process is reviewed. This method uses the Furry Picture and the exact solutions of the Dirac equation in the external field potential. For future linear colliders, the nominal machine parameters are such that this external field can be considered to be a constant crossed electromagnetic field. The Dirac equation solutions can be Fourier transformed such that they are functions of Dirac gamma matrices, Airy functions and the usual non-external field solution. The resultant analytic form for the transition rate is the same as that obtained by the Quasiclassical Operator (QO) method of Baier-Katkov which is valid in the limit of  ultra-relativistic electron and vanishingly small radiation angle. The NR calculation however also exhibits a pole in the radiation angle for back-radiated photons. The removal of this pole requires a further study of IR divergences within the Furry Picture.
\end{abstract}

\section{Introduction}
There has been a revival in interest over the past decade or so, in fundamental QED processes occurring in intense external fields. A SLAC experiment involving the interaction of an intense laser with an electron beam showed clearly the onset of multiple numbers of laser photons contributing to the process \cite{Bamber99}. Around the same time a similar set of experiments in Europe, on electron laser interaction were performed which observed a fermion mass shift due to the action of the electron motion in the laser field \cite{Meyerhofer95,Meyerhofer96}. Another set of experiments has taken place at CERN during the last few years involving fermion processes in the intense fields present in the atomic lattices of crystals \cite{Uggerhoj05}. Good agreement has been found with theoretical calculations of QED process in crystals \cite{Baier83}.\\

The ratio of the strength of the external field to the Schwinger critical field (at which real $e^+e^-$ pairs can be produced from the vacuum) gives a measure as to whether a quantum or classical treatment of the interaction with the external field is necessary. An 'intense field', then, has a field strength sufficiently high to require a quantum treatment. Several calculation strategies exist.\\

 If the field strength is not too high the external field can be considered to contribute one or more photons in the Born approximation \cite{Schroeder}. Higher field strengths require that collective effects be taken into account  by performing a QED calculation in the Furry Picture (FP). In this picture the external field potential is bundled up with the Dirac field to form a bound Dirac field which interacts with the free Maxwell field. Exact solutions of the Dirac equation in a (classical) external potential exist and are used to calculate the transition probabilities of interest. \\

Two calculation strategies are compared in this paper. In the first strategy (also first historically) the transition rates of various first order processes in the FP were calculated using the usual S-matrix theory. The interpretation of the external field as photons contributing to the process emerges naturally. Exact results for transition rates with no kinematic approximations can be obtained.  This method is referred to as the Nikishov-Ritus (NR) method in this paper \cite{NikRit64a,NikRit64b}.\\

The second method makes use of the fact that in particle colliders the initial state fermions are ultra-relativistic and the energy levels of the fermion states in the external field of nominally designed bunches are extremely close together. If the fermion final states are also ultra-relativistic then the fermion motion can be considered classical in the external field. This Quasi-classical Operator (QO) method starts from an intermediate point in perturbation theory before the integration with respect to time has been carried out. The transition probability is obtained after allowing the operators of the electron motion to commute \cite{Baier67}.\\

The first order processes in the FP for a number of different external fields, including electromagnetic fields of various polarisations, magnetic fields and intra-crystalline fields, can be considered well studied. Second order processes in the FP require substantial further study, as do the radiative corrections \cite{Hartin06}. In this paper the beamsstrahlung process, i.e the bremsstrahlung in the electromagnetic field of a relativistic particle bunch, is considered. The NR method for this process will be reviewed and the transition rate will be compared to that obtained using the QO method.\\

One application of interest for this work is that of spin tracking through the beam-beam interaction at colliders. The CAIN program \cite{Cain} which performs this spin tracking, assumes no radiation angle for the outgoing photon. In the interests of precision spin tracking, especially given that collider designs include a crossing angle, it is important to know the variation of calculated transition probability with radiation angle and no kinematic approximations \cite{Bailey06}. The full NR result will therefore be examined for this behaviour.

\section{QED in intense fields}
The motivation for using the FP arose in the studies of electrons in atomic systems in which bound states were natural ones. The interaction of the electrons with the external field can be handled by iteration using the Born approximation. However for strong enough fields the Born series does not converge and the interaction with the external field has to be taken into account to all orders. The FP achieves this by repackaging the Lagrangian density of the system with the interaction term specifying the coupling between free Maxwell and the combined Dirac-external field system. The usual perturbation method within S-matrix theory is then employed using the solution for the bound Dirac field equation, 

\smallskip\begin{equation}\label{c2.eq21}\begin{array}{c}
\left[ (p-eA^e)^2-m^2-\frac{ie}2F^{e}_{\mu \nu }\sigma^{\mu \nu}\right] \psi _V(x,p)=0 \\[20pt]
\text{where} \quad\quad F^{e}_{\mu \nu }=k_\mu \dfrac{\partial A^e_\nu }{\partial \phi }-k_\nu 
\dfrac{\partial A^e_\mu }{\partial \phi } \quad\quad\text{and}\quad\quad \phi =k^\mu x_\mu
\end{array}\end{equation} \medskip

The solution $\psi _V(x,p)$  to the bound Dirac field equation  was worked out long ago \cite{Volkov35}. A nice exposition of its derivation is found in section 40 of \cite{BPLqed}. This Volkov solution is a product of a Volkov $E_p(x)$ function of the external field 4-potential $A^e_\mu$, and the usual free field solution,

\begin{equation}\label{c2.eq24}\begin{array}{c}
\psi _V(x,p) = E_p(x)\;\exp(-ip\cdot x)\;u_s(p) \\[10pt] 
E_p(x) = \left[ 1+\dfrac e{2(k\cdot p)}\slashed{k}\slashed{A}^e\right] \exp \left(
-i\dint_{\nths 0}^{kx}\left[ \dfrac{e(A^e\cdot p)}{(k\cdot p)}-\dfrac{e^2A^{e2}}{2(k\cdot p)} \right] d\phi \right) 
\end{array}\end{equation}\medskip

The orthogonality of the Volkov $E_p(x)$ is an important result \cite{Guo91,Zako05} which allows the bound fermion propagator $G^e(x,x^{\prime })$ to be written as the usual propagator sandwiched between Volkov $E_p(x)$ functions,

\begin{equation}\label{c2.eq26}
G^e(x,x^{\prime })=\frac 1{(2\pi )^4}\int_{\nths -\infty}^\infty d^4p \,E_p(x)
\frac{\slashed{p}+m}{p^2-m^2}\bar E_p(x^{\prime }) 
\end{equation} \medskip

Using the above results the required interaction between bound fermions and photons can be studied using the usual S-matrix theory and Wick's theorem. Since the Volkov solution is written in terms of the bispinor and gamma matrices the usual Dirac gamma algebra and spin and polarization sums are used to obtain the transition probability of the desired QED process. This was the route chosen by Nikishov and Ritus (and others) in the 1960s and which is summarised in the following two sections.\\

With the solutions of the bound Dirac equation and an expression for the bound fermion propagator, Volkov $E_p(x)$ functions can be grouped together to form only one new element in the S-matrix - a modified vertex 

\begin{equation}\label{eqn5} 
\Gamma^e_\mu=\int d^4x\;\bar{E}_{p_f}(x) \gamma_\mu E_{p_i}(x) \exp\left[ ix\cdot (p_i-p_f-k_f) \right] 
\end{equation}\medskip

In terms of the Feynman diagrams, the bound Dirac equation solutions are represented as double straight lines. 

\section{Bremsstrahlung in a constant crossed field}

\begin{figure}[Htb]
\begin{minipage}{0.4\linewidth}
\begin{center}
 \includegraphics[width=0.75\linewidth]{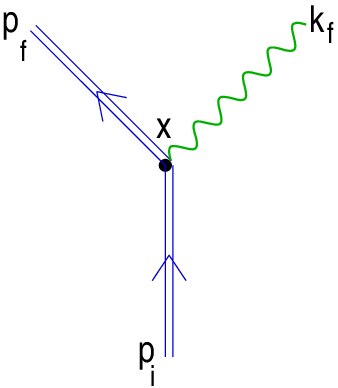}
\caption{\label{fig1}Beamsstrahlung Feynman diagram.}
\end{center}
\end{minipage}\hspace{0.15\linewidth}
\begin{minipage}{0.4\linewidth}\centering
\begin{center}
 \includegraphics[width=0.8\linewidth]{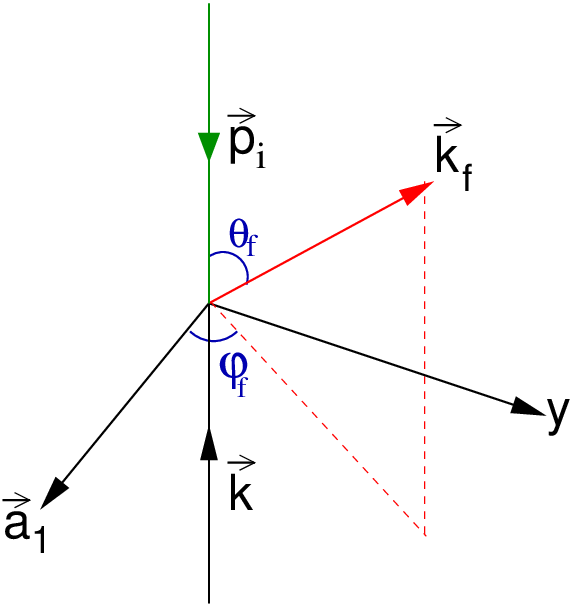}
\caption{\label{fig2}Coordinate system and radiation angles.}
\end{center}
\end{minipage} 
\end{figure}

The beamsstrahlung process (figure \ref{fig1}) is simply the bremsstrahlung process that takes place when a charge radiates a photon in the field of an oncoming charge bunch in collider collisions. For proposed future linear colliders the radiation coherence length - classically, the ratio of the radius of curvature of the radiating charge to the relativistic gamma value of the radiating charge - is shorter than both the typical bunch length and the characteristic length specified by the beam disruption parameter. Therefore the external field can be considered to be a constant crossed field \cite{YokChen91} specified by

\begin{equation}\label{eqn6}\begin{array}{c} 
A^e_\mu=(k\cdot x)\;a_{1\mu}\\[8pt]
(a_1\cdot a_1)=-a^2
\end{array}\end{equation}\medskip

The Lorentz condition and a choice of reference frame such that the zeroth (the time) component of the external field 4-potential is zero, requires that the 3-potential $\vec{a}_1$ and the 3-momentum $\vec{k}$ of the external field  be orthogonal. Using these vectors as a basis for a coordinate system with $\vec{a}_1$ forming the x-axis and $\vec{k}$ the z-axis, the radiation angles $(\theta_f,\phi_f)$ of the radiated photon $\vec{k}_f$ can be specified (figure \ref{fig2}). If the radiating electron is directed exactly opposite to the wave-vector of the oncoming bunch field (as is usually the case in colliders except for a small crossing angle) then the angle $\theta_f=\pi$ corresponds to forward radiation and $\theta_f=0$ to back radiation \\

An explicit form for the Volkov $E_p(x)$ functions can be obtained by substitution of the 4-potential of the constant crossed field. In order to simplify the dependence on the space-time variable $x_\mu$, a Fourier transform is made to an infinite integration over a variable $r$

\begin{equation}\label{eqn7}\begin{array}{c} 
(k\cdot x)^n \exp(i S(x))=\dint_{\nths -\infty}^\infty dr \; F_n(r)\exp(-ir(k\cdot x)) \\[10pt]
\text{where} \quad F_n(r)=\dint_{\nths -\infty}^{\infty} dt \; t^n\exp(irt+iS(t)) \\[10pt]
\text{and} \quad S(t)=-\left[ \dfrac{e(a_1\cdot p)}{2(k\cdot p)}t^2+\dfrac{e^2a^2}{6(k\cdot p)}t^3 \right]
\end{array}\end{equation}\medskip

The modified vertex can likewise be transformed and the integration over space-time carried out to  produce the usual delta function whose argument expresses the momentum conservation 

\begin{equation}\label{eqn8}\begin{array}{c} 
\Gamma^e_\mu(r)=(2\pi)^4 \dint _{\nths -\infty}^{\infty} dr\; \bar E_{p_f}(r) \gamma_\mu E_{p_i}(r) \;\delta^4(p_f+k_f-p_i-rk)
\end{array}\end{equation}\medskip

There is a contribution $rk_\mu$ from the external field, which is integrated over all values allowed by conservation of momentum. It is open to interpretation whether this is indeed one photon of 4-momentum $rk_\mu$ or $r$ photons of 4-momenta $k_\mu$ each. This is unlike the case for processes in laser fields in which the contribution from the external field is a summation (over, say, $n$) rather than an integration and the interpretation clearly is in terms of $n$ external field photons. Though external field photons were not included at the outset, they emerge through the quantum interaction of the electron with the external field.\\

For a constant crossed field, the concept of a quasi-momentum for the fermions $q_\mu=p_\mu+\frac{e^2a^2}{2(k\cdot p)}\;k_\mu$ is ill defined, since the intensity of the crossed field $\frac{e^2a^2}{m}\rightarrow\infty$. Consequently the conservation of momentum of the process is written in terms of the usual external field-free momentum $p_{\mu}$.

\section{Beamsstrahlung transition probability}
The matrix element of the beamsstrahlung process is the modified vertex sandwiched between Dirac bispinors. In deriving the differential transition probability, $dw$, it proved useful to use relativistic normalization for the field operators and to follow the derivation of the general form in \cite{PesSchr95}. It should be noted that the square of the matrix element will introduce another integration (over variable $r'$) and a second delta function. The particle energy-momenta are specified by $(\epsilon_i,\vec{p}_i)$, $(\epsilon_f,\vec{p}_f)$ and $(\omega_f,\vec{k}_f)$.

\begin{equation}\label{eqn9}
dw=\frac{1}{16\epsilon_i\pi^2} \overline{\sum_{if}} \left|\overline{u}(p_f)\Gamma^e_\mu(r)u(p_i) \right|^2 \; \frac{d^3\vec{p}_f}{\epsilon_f} \frac{d^3 \vec{k}_f}{\omega_f}
\end{equation}\medskip

There is nothing conceptually difficult in reducing the transition probability to its simplest form, however the procedure is technical and time-consuming. The main steps are simply listed and the end result quoted for final analysis:\bigskip
\begin{itemize}
 \item Perform the integration over initial 3-momenta using 3 components of one of the delta functions
 \item Write the Fourier transformed functions $F_n$ as Airy functions by using the relation 
 \[ \dint^\infty_{\nths -\infty}\;dt\;\exp\left(iat+bt^2+i\frac{c}{3}t^3\right)=c^{-1/3} \text{Ai}\left[c^{-1/3}\left(a-\frac{b^2}{c}\right)\right] \exp\left[i\frac{b}{c}\left(\frac{2b^2}{3c}-a\right)\right] \]
  \item Perform the spin sums and trace calculation using the usual Dirac algebra
  \item Introduce an additional delta function using the relation 
  \[ \int\;\frac{d^3\vec{p}_f}{2\epsilon_f}=\dint\;d^4p_f \;\delta(p_f^2-m^2) \] 
  \item Using two delta functions do the integrations over final fermion 3-momenta $\vec{p}_f$ and the contribution from the external field $r$, to get the condition $r\rightarrow (p_i\cdot k_f)/(k\cdot p_f)$ \vspace{6pt}
   \item Integrate over the x-component of the final photon momentum $k^x_f$ to get the condition that $r'\rightarrow r$ and transform the transition probability $w$ to a transition rate $W$ \vspace{6pt}
   \item Transform the y-component of the final photon 3-momentum $k^y_f$ and integrate to reduce products of Airy functions to a single Airy function \cite{Aspnes} using
 \[ \dint_{\nths 0}^{\infty} \frac{dt}{\sqrt{t}}\text{Ai}^2(t+a)=\frac{1}{2}\dint^\infty_{\nths 2^{2/3}a} \text{Ai}(t)\; dt \]
   \item Replace Airy functions with Bessel functions using relations like,
$ \text{Ai}(t)=\frac{\sqrt{t}}{\pi\sqrt{3}} \text{K}_{1/3}\left(\frac{2}{3}t^{3/2}\right) $ \vspace{6pt}
    \item The  integration over the z-component of the final photon 3-momenta  $k^z_f$ is transformed to a function of scalar products of 4-momenta $u=\frac{(k\cdot k_f)}{(k\cdot p_f)} $
\end{itemize}\bigskip

The final result for the transition rate is

\begin{equation}\label{eqn10}\begin{array}{c}
 \dfrac{dW}{du} =\dfrac{\alpha m^2}{\pi\sqrt{3}\epsilon_i} \dfrac{1}{(1+u)^2}\left[\dint_{\nths\chi}^\infty K_{5/3}(y)dy
     -\dfrac{u^2}{1+u}K_{2/3}(\chi)\right] \\[20pt]
\text{where}\quad\quad  \chi=\dfrac{2u}{3ea(k\cdot p_i)} \quad \text{and} \quad u=\dfrac{(k\cdot k_f)}{(k\cdot p_i)-(k\cdot k_f)} 
\end{array}\end{equation}\medskip

\section{Comparison of the Nikishov-Ritus and Baier-Katkov methods}
The same beamsstrahlung transition rate appears in the literature for both NR and QO methods, though the exact form of the integration variable varies from derivation to derivation and NR prefers to write in terms of Airy functions. Nevertheless, a direct comparison can be made and the difference in the final result for the two methods is encapsulated in the integration variable $\chi$, 

\begin{equation}\label{eqn10b}\begin{array}{c}
\chi(\text{QO})=\dfrac{1}{3ea(k\cdot p_i)}\dfrac{\omega_f}{\epsilon_i-\omega_f}  \\[20pt]
\chi(\text{NR})=\dfrac{2}{3ea(k\cdot p_i)}\dfrac{(k\cdot k_f)}{(k\cdot p_i)-(k\cdot k_f)} 
\end{array}\end{equation}\medskip

Physically, the parameter $\chi$ is inversely proportional to the magnetic field strength of the oncoming charge bunch and proportional to the ratio of radiated photon energy to final electron energy. The $\chi(\text{NR})$ also retains an angle dependence in its scalar products. Naively, if the electron is ultra-relativistic  and radiates almost entirely in the forward direction $\theta_f=\pi$ (as one might expect) then $\chi(\text{NR})\rightarrow\chi(\text{QO})$ and the two results seem consistent in the limit assumed by the QO method.\\

However, the beamsstrahlung transition rate contains a pole at the lower end of the $\chi$ integration (at $\chi=0$) since the $K$ Bessel function of fractional order approaches infinity as its argument approaches zero \cite{AbrSte}. Comparison between the QO and NR methods near the pole requires careful consideration. Both $\chi(\text{NR})$ and $\chi(\text{QO})$ tend towards zero as the energy of the radiated photon becomes  small. This is the IR divergence. Applying the usual argument for the removal of the divergence, a cut is made at a minimum energy which is the limit of the energy resolution of the detector. The residual divergent pole is then removed theoretically by inclusion of radiative corrections, though it should be noted that it does not appear to have been done explicitly in the literature for this process in the FP.\\

\begin{figure}[Ht!]
\centering
\subfigure{ \includegraphics[scale=0.4]{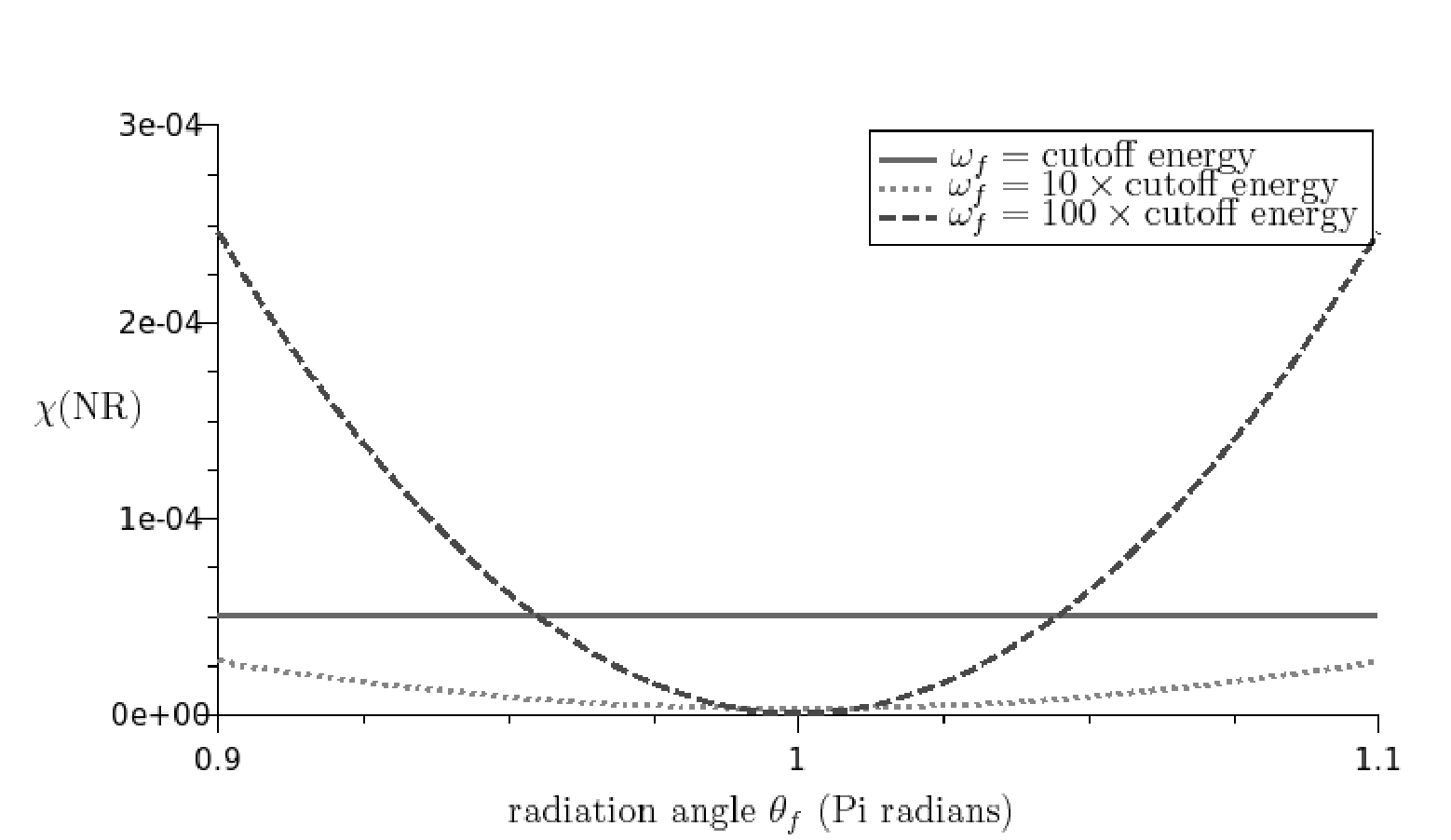}
\label{fig3}}
\subfigure{\includegraphics[scale=0.4]{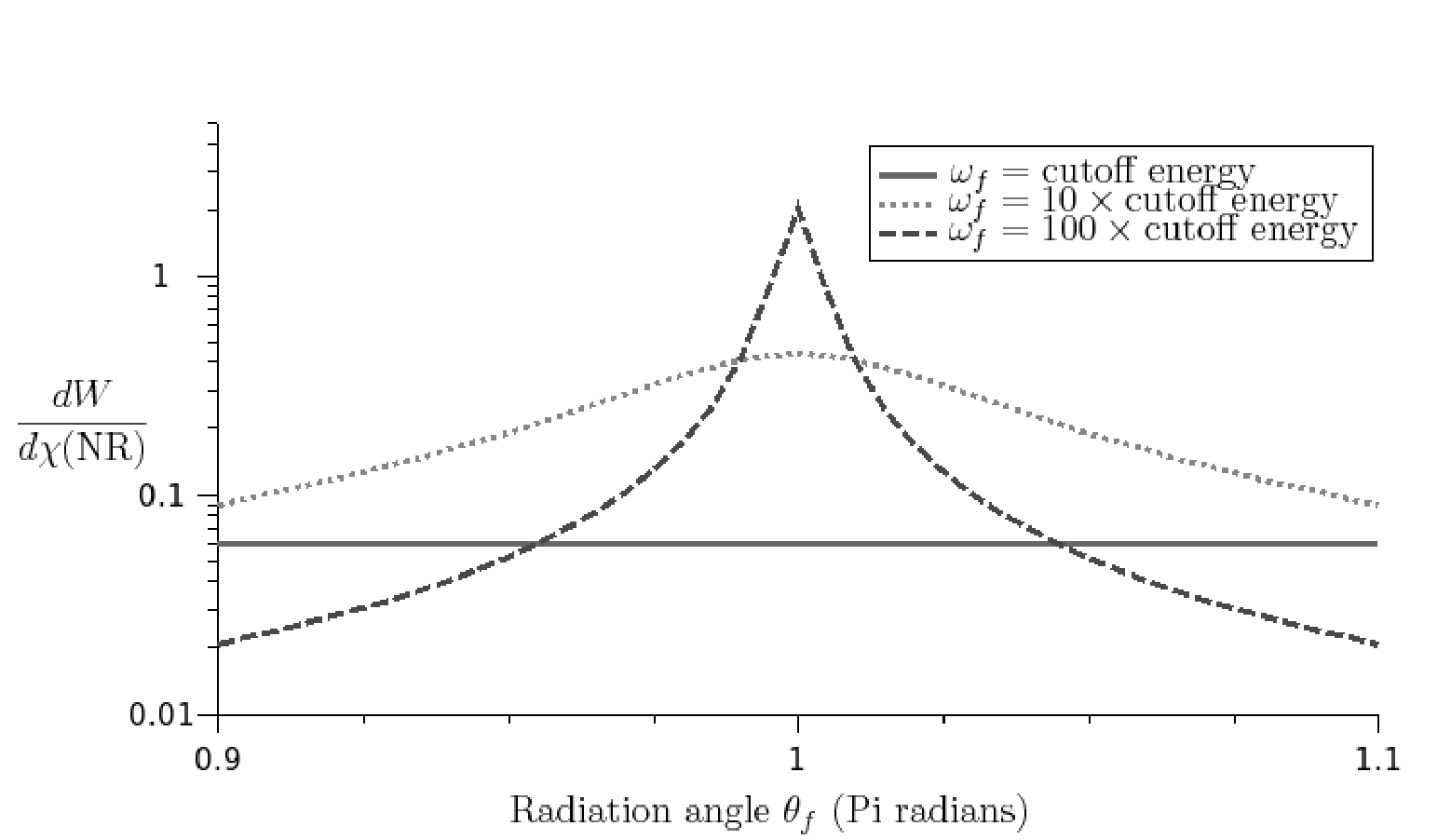}
\label{fig4}}
\caption{Differential transition rate $\dfrac{dW}{d\chi\text{(NR)}}$ and parameter $\chi$(NR) near the transition rate pole.}
\end{figure}

Writing the minimum photon energy cut instead as a fictious small photon mass $m_\gamma$ the parameter $\chi(\text{NR})$ can be written in terms of the radiation angle \\

\begin{equation}\label{eqn10c}\begin{array}{c}
\chi(\text{NR})=\dfrac{2}{3ea(k\cdot p_i)}\;\frac{\sqrt{|\vec{k}_f|^2+m_\gamma^2}-|\vec{k}_f|\cos(\theta_f)}
{2\epsilon_i-\sqrt{|\vec{k}_f|^2+m_\gamma^2}-|\vec{k}_f|\cos(\theta_f)} 
\end{array}\end{equation}\medskip

and its dependence plotted (figure \ref{fig3}). The closer $\chi(\text{NR})$  reaches zero, the more closely the transition rate pole is reached. Clearly this is the case for $\theta_f=0$, i.e. back radiation and is the case even for ultra-relativistic electrons. This is not a trivial observation since even with an energy cut, the transition rate reaches a maximum around the pole and any small variation at that point will have a relatively large effect on the transition rate value.\\

 A maximum transition rate for back radiation is obviously in contradiction to the results of classical electrodynamics which predict forward radiation for ultra relativistic electrons. Additionally the pole in the transition rate for back radiation renders the equivalence of the QO and NR results within the ultra relativistic limit invalid.\\

In principle, the removal of the IR divergence from the beamsstrahlung process within the FP should be straightforward. It has been argued that since IR divergences cancel in the normal interaction picture, then it must also be the case for the bound interaction picture \cite{Rohrlich55}. The removal may require a calculation of the vertex function in the FP. In any case a dedicated further study is necessary.

\section{Conclusion}

The analytic form of the transition rate of the beamsstrahlung process has been compared for two calculation methods. The QO method assumes a classical interaction between the electron and the external field, but a quantum interaction to produce the radiated photon. The spectrum of the radiation (the derivative of the transition rate with respect to the radiated photon energy) shows the usual IR divergence for low energy photons. The QO calculation is only expected to be valid for an ultra-relativistic electron before and after the radiation.\\

The NR method is a full calculation using the bound interaction picture within the usual S-matrix theory with no kinematic approximations. In certain limits the analytic form of the NR radiation intensity appears to reduce to that of the classical Schott formula \cite{NikRit67} and to the beamsstrahlung transition rate obtained by the QO method. However a full consideration of the NR calculation reveals a transition rate maximum for back-radiated photons. It seems clear that an explicit inclusion of radiative corrections within the FP is required and such a work is intended for a forthcoming paper.\\

\ack This work has been supported by the DFG via the Emmy-Noether grant Li 1560/1-1.

\end{document}